\documentclass[aps,prl,reprint]{revtex4-1}

\usepackage{graphicx}
\usepackage{amssymb}
\usepackage{amsmath}
\usepackage{color}
\usepackage{paralist}
\usepackage{bm}
\usepackage{verbatim}

\usepackage{amsfonts}
\usepackage{epstopdf}

\def\red{\color{red}}

\begin{document}
%%%%%%%%%%%%%
\title{Observation of sub-kelvin superconductivity in Cd$_3$As$_2$ thin films}

\author{
A.V. Suslov$^{\,1}$, A.B. Davydov$^{\,2}$, L.N. Oveshnikov$^{\,3,2,*}$, L.A. Morgun$^{\,2,4}$, K.I. Kugel$^{\,4,5}$, V.S. Zakhvalinskii$^{\,6}$, E.A. Pilyuk$^{\,6}$, A.V. Kochura$^{\,7}$, A.P. Kuzmenko$^{\,7}$, V.M. Pudalov$^{\,2,5}$, B.A. Aronzon$^{\,3,2}$
}
\affiliation{
$^{1}$National High Magnetic Field Laboratory, Tallahassee, Florida, 32310 USA
\\
$^{2}$P.~N.~Lebedev Physical Institute, Russian Academy of Sciences, Moscow, 119991 Russia
\\
$^{3}$National Research Center ``Kurchatov Institute", Moscow, 123182 Russia
\\
$^{4}$Institute for Theoretical and  Applied Electrodynamics, Russian Academy of Sciences, Moscow, 125412 Russia
\\
$^{5}$National Research University Higher School of Economics, Moscow, 101000 Russia
\\
$^{6}$Belgorod National Research University, Belgorod, 308015 Russia
\\
$^{7}$South-West State University, Kursk, 305040 Russia
}

%%%%%%%%%%%%%%%%%%%%%%%%%%ABSTRACT%%%%%%%%%%%%%%%%%%%%%%%%%%%%%%%%%%%%%%%%%%%%%%%%%%%%%%%%%%%%%%%%%%%%%%%%%%%%
\begin{abstract}
We report the first experimental observation of superconductivity in Cd$_3$As$_2$ thin films without application of external pressure. Surface studies suggest that the observed transport characteristics are related to the polycrystalline continuous part of investigated films with homogeneous distribution of elements and the Cd-to-As ratio close to stoichiometric Cd$_3$As$_2$. The latter is also supported by Raman spectra of the studied films, which are similar to those of Cd$_3$As$_2$ single crystals. The formation of superconducting phase in films under study is confirmed by the characteristic behavior of temperature and magnetic field dependence of samples resistances, as well as by the presence of pronounced zero-resistance plateaux in $dV/dI$ characteristics. The corresponding $H_c-T_c$ plots reveal a clearly pronounced linear behavior within the intermediate temperature range, similar to that observed for bulk Cd$_3$As$_2$ and Bi$_2$Se$_3$ films under pressure, suggesting the possibility of nontrivial pairing in the films under investigation. We discuss a possible role of sample inhomogeneities and crystal strains in the observed phenomena.
\end{abstract}
%%%%%%%%%%%%%%%%%%%%%%%%%%%%%%%%%%%%%%%%%%%%%%%%%%%%%%%%%%%%%%%%%%%%%%%%%%%%%%%%%%%%%%%%%%%%%%%%%%%%%%%%%%%%%%%%

%\pacs{72.10.-d, 72.15.Lh, 72.80.Vp, 72.80.Ng}

%%%%%%%%%%%%%%%%%%%%%%%%%%%%%%%%%%%%%%%%%%%%%%%%%%%%%%%%%%
%71.55.Ak	Metals, semimetals, and alloys
%72.10.-d	Theory of electronic transport; scattering mechanisms
%78.40.Kc	Metals, semimetals, and alloys
%72.15.Lh	Relaxation times and mean free paths (in metals and alloys)
%72.80.Ng	Disordered solids (under transport in specific materials)
%72.80.Vp	Electronic transport in graphene

\date{\today}

\maketitle

%%%%%%%%%%%%%%%%%%%%%%%%%%%%%%%%%%%%%%%%%%%%%%%%%%%%%%%%%%%%%%%%%%%%%%%%%%%%%%%%%%%%%

Weyl and Dirac semimetals (WSM and DSM) currently attract wide interest related to the existence of Dirac nodes in their electron spectrum and related nontrivial topological characteristics of both bulk and surface states~\cite{ArmitageRMP2018,YanAnnRevCMPh2017,WangAdvPhX2017}. A special attention is drawn to the Cd$_3$As$_2$ compound, which proved to be air-stable, unlike some other DSM materials~\cite{Neupane_NatCom2013,Liu_NatMat2014,Jeon_NatMat2014, Borisenko_PRL2014,LuScience2015}. This compound is known and has been studied for quite a long time~\cite{DubowskiAPL1984,WeclewiczTSF1987,ShazlyVac1996,JarzabekJNCS2004}. Nevertheless, it is still very popular since the Dirac nodes of this semimetal are protected by the crystal symmetry and the electron states  exhibit interesting topological properties, such as spin-momentum locking. Due to the high symmetry, DSM materials can undergo transitions to other topological phases (e.g. WSM) as a result of breaking certain symmetries or applying some external factors.

The existence of topologically protected electron states gives a new impetus to the problem of topological superconductivity (TSC) being discussed for a long time~\cite{WrayNatPh2010,AndoAnnRevCMPh2015,SatoRPP2017}. In particular,  Refs.~\cite{SatoPRL2015,SatoPRB2016} provide a theoretical analysis of the possible types of SC pairing in the systems of Cd$_3$As$_2$ type, including topologically nontrivial ones. Basically, the existence of nontrivial pairing potential (leading to the emergence of triplet Cooper pairs) should induce the formation of Majoranna modes at the surface of the crystal, which can be used in the fault-tolerant quantum computing~\cite{AndoAnnRevCMPh2015}. The SC phase emergence in Cd$_3$As$_2$ was reported in Ref.~\cite{HeNPJQuMat2016}, however, it was observed at pressures above the structural transition from the tetragonal to trigonal phase. While theoretical works suggest the stabilization of TSC phase upon such symmetry lowering~\cite{SatoPRL2015,SatoPRB2016}, such transition also implies appearance of the gap at Dirac nodes, thus, suppressing the DSM phase. Additional indications of the SC phase in Cd$_3$As$_2$ were also observed in the point-contact spectroscopy experiments and attributed to the local crystal distortions under point-contact~\cite{WangNatMat2016,AggarwalNatMat2016}. It is important to note, that up to now, all indications of the SC phase emergence in Cd$_3$As$_2$ were related either to the pressure-induced structural changes or to the proximity effect with the conventional \emph{s}-wave SC~\cite{LiPRB2018,YuPRL2018}.

Recent studies of Cd$_3$As$_2$ thin films outlined their specific features. The observation of quantum Hall state~\cite{Uchida2017Natcomm,Schumann2018PRL} and consecutive analysis of magnetoresistance of Cd$_3$As$_2$ films suggest the existence of closed Fermi loops (Weyl orbits) related to the surface states, below some critical thickness~\cite{Zhang2017NatComm,gall2018}. Thus, superconducting Cd$_3$As$_2$ thin film with nontrivial pairing (TSC state) should yield surface Majoranna modes, which, however, can be modified (in comparison to the bulk crystal~\cite{SatoPRL2015}) due to interaction of opposite surfaces.

In the previous studies bulk Cd$_3$As$_2$ has not exhibited any indications of superconductivity except that arising at extremely high pressure. In contrast to this, we have observed superconductivity in Cd$_3$As$_2$ thin films without external compression. We argue that the observed phenomena cannot be related to any parasitic effects and it have certain similarities to the previously observed SC in Cd$_3$As$_2$ under pressure~\cite{HeNPJQuMat2016}, suggesting the possible presence of nontrivial pairing in the studied films.

The films under study were deposited by magnetron sputtering. Single crystals of Cd$_3$As$_2$ used for sputtering were grown by vapor phase deposition from high-purity initial components. In this paper, we present the results for three films, deposited on Si and Al$_2$O$_3$ substrates and shaped into Hall bar geometry with a mm-scale conduction channel.

The initial surface studies of the grown films were conducted using the scanning electron microscope (SEM) JSM-6610LV (Jeol, Japan) with the additional X-Max$^N$ module (Oxford Instruments, UK) for energy-dispersive X-ray spectroscopy (EDXS). For imaging, we used detectors for secondary and backscattered electrons. The additional studies were performed using atomic-force microscope (AFM) SmartSPM 1000 (AIST NT, USA). Raman spectra were recorded at room temperature using combined scanning probe microscopy system with the confocal fluorescence spectrometer and Raman spectrometer OmegaScope$^{\mathrm{TM}}$ (AIST NT, USA). For excitation we used a laser with 532 nm wavelength, power of 50 mW, and the focused light spot at the sample surface of about 500 nm. The spectral resolution was 0.8 cm$^{-1}$.

To eliminate all uncertainties related to the experimental artefacts in measurements of the transport properties of the samples, we used two different setups. Samples A and B were studied using a cryogen-free dilution refrigerator BF-LD250 with a 1 Tesla magnet (BlueFors, Finland) at the Lebedev Shared Facility Center, Moscow, Russian Federation. Those experiments were performed by the conventional low-frequency (7.142 Hz) lock-in four-probe method with measurement currents (of the order of 100 nA) considerably lower than critical values obtained for studied films. Sample C was studied using a 20 Tesla superconducting magnet with a dilution refrigerator (SCM1) at the National High Magnetic Field Laboratory, Tallahassee, Florida, USA. There a Model 372 ac resistance bridge with a preamp 3708 (Lake Shore Cryotronics, Inc., USA) was used for magnetoresistance testing by the four-probe method at current 316 nA. Measurements of the differential resistance dV/dI were performed utilizing a current source Model 6221 and a nanovoltmeter Model 2182A (Tektronix, Inc., USA).

SEM images of the studied films reveal similar surface morphology. The EDX mapping reveals a homogeneous distribution of components, implying that transport characteristics of the film are related solely to the Cd-As binary system. The EDXS demonstrates that at $\mu$m scale, the Cd-to-As ratio is close to stoichiometric Cd$_3$As$_2$ within the 2$\%$ accuracy. The presence of Cd$_3$As$_2$ phase is also supported by the Raman spectroscopy results. A typical Raman spectrum for studied films reveal the presence of two distinct peaks similar to those observed for Cd$_3$As$_2$ nanocrystallites~\cite{Wei2006CGD} and thin films~\cite{Cheng2016NJP}. Surface morphology of studied samples determined via AFM suggests the polycrystalline structure of the bulk of films under investigation. The thickness of studied films determined by AFM is 40{\red --}50 nm for samples A and C, and about 80 nm for sample B.

\begin{figure}[t]
\begin{center}
  \includegraphics[width=1\columnwidth]{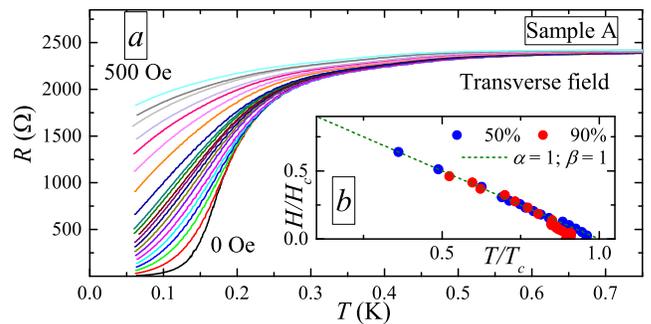}
\end{center}
   \caption{\label{S38}
(a) Temperature dependence of resistance for sample A at various transverse magnetic fields. (b) Corresponding $H_c-T_c$ diagrams for SC transition estimated as a resistance drop to 50$\%$ (midpoint) and 90$\%$ of the normal value. Fitting of experimental data by Eq.~(\ref{crit}) is shown by the dashed line.
 }
\end{figure}

\begin{figure}[t]
\begin{center}
  \includegraphics[width=1\columnwidth]{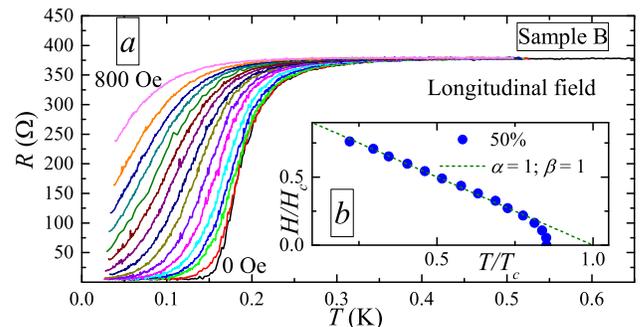}
\end{center}
   \caption{\label{S58}
(a) Temperature dependence of resistance for sample B at various longitudinal magnetic fields. (b) Corresponding $H_c-T_c$ diagrams for SC transition (midpoint). Fitting of experimental data by Eq.~(\ref{crit}) is shown by the dashed line.
 }
\end{figure}

Transport measurements of the studied films reveal the clearly pronounced SC transition below 0.5\,K. In Fig.~\ref{S38}a, we show the  temperature dependence of resistance for sample A at various values of the transverse magnetic field. As one can see, there is a distinct resistance drop (to almost zero value) that shifts to the lower temperatures upon increasing magnetic field, which supports the assumption of SC phase emergence. Sample B demonstrates analogous behavior (see Fig. \ref{S58}a). The actual transition regions are rather broad, which agrees well with the polycrystalline structure of the films.

\begin{figure}[t]
\begin{center}
  \includegraphics[width=1\columnwidth]{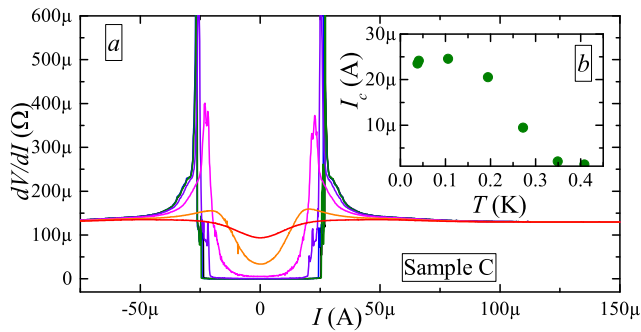}
  \end{center}
   \caption{\label{S16a}
(a) Differential resistance for sample C at various temperatures. (b) Corresponding temperature dependence of critical current, $I_c$.
 }
\end{figure}

\begin{figure}[t]
	\begin{center}
		\includegraphics[width=1\columnwidth]{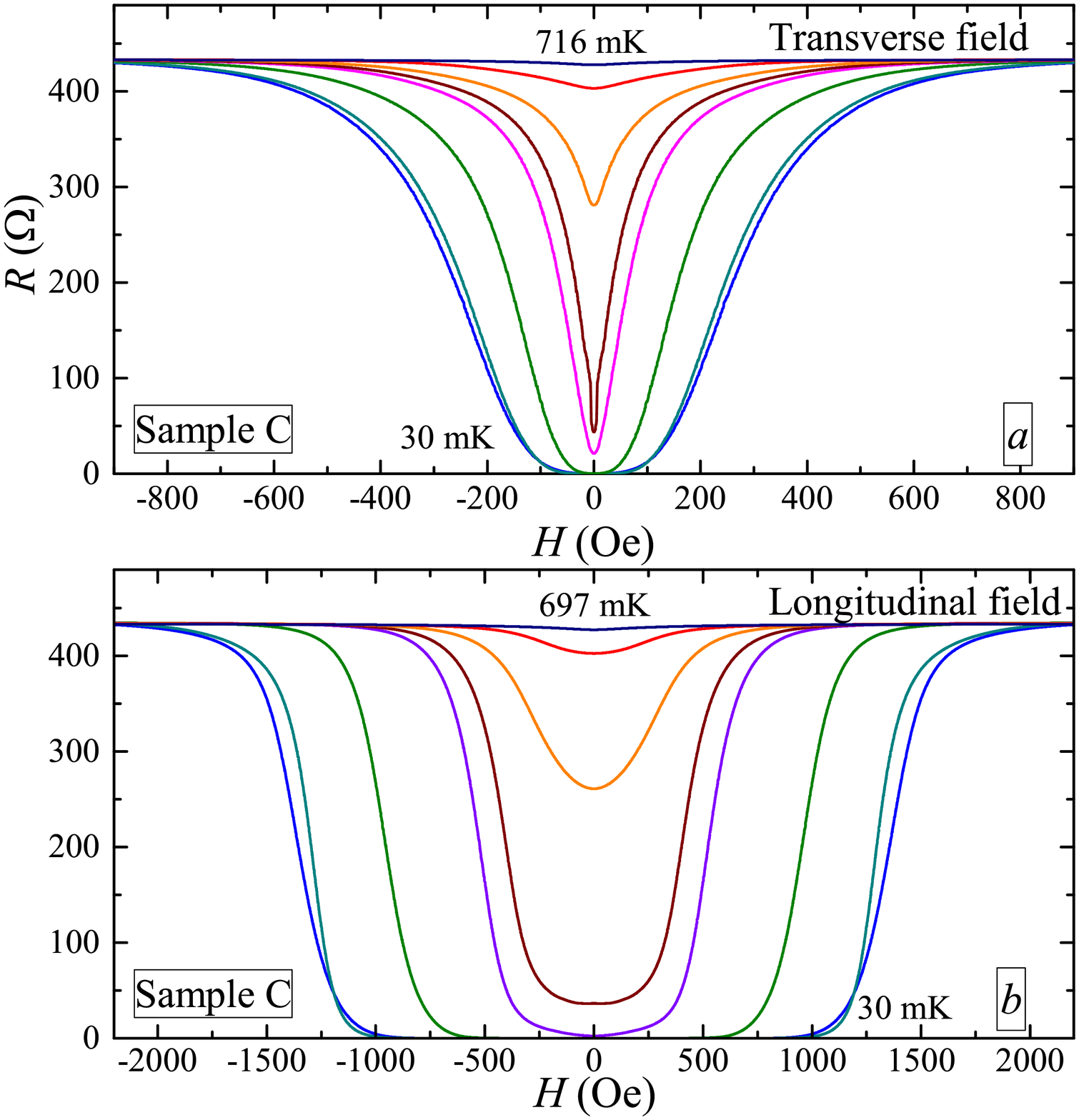}
		\includegraphics[width=1\columnwidth]{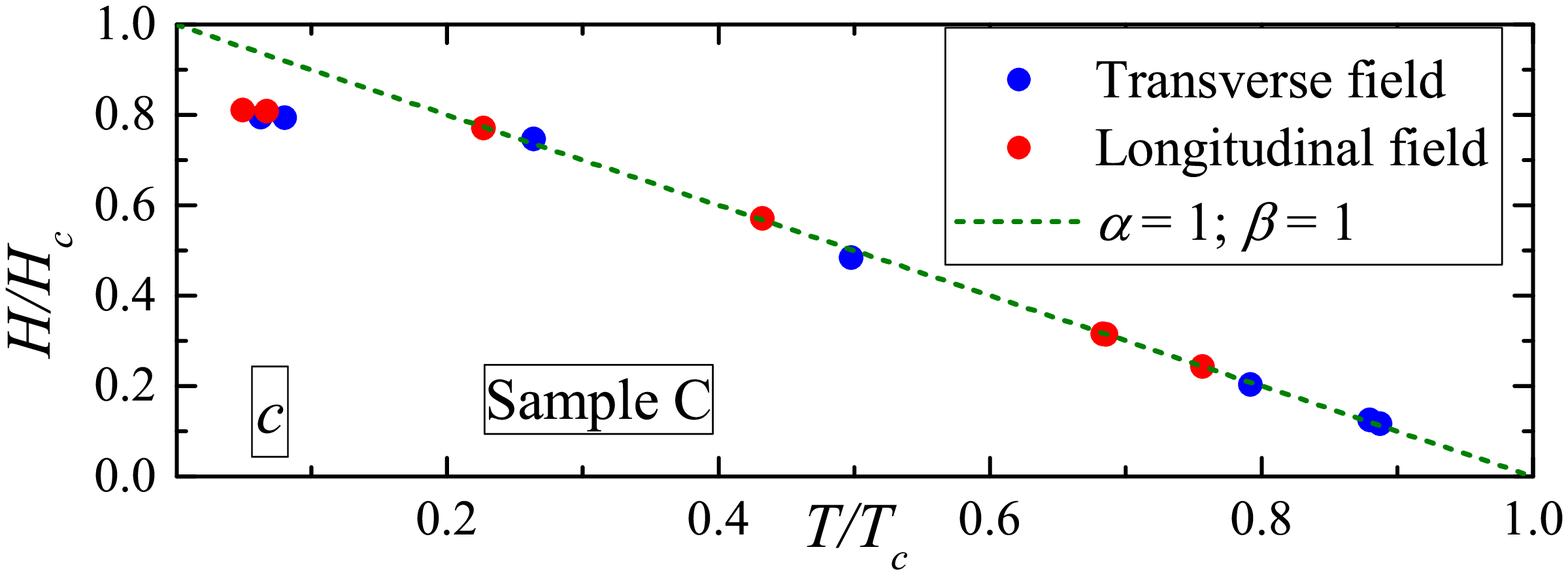}
	\end{center}
	\caption{\label{S16b}
Low-field magnetoresistance of sample C at various temperatures at (a) transverse and (b) longitudinal magnetic fields. (c) Corresponding $H_c-T_c$ diagrams for the SC transition.
	}
\end{figure}

The $dV/dI$ characteristics for sample C measured at various temperatures clearly demonstrate zero-resistance plateaux (Fig.~\ref{S16a}a). The critical current values, $I_c$, decrease as the temperature approaches $T_c$ (Fig. \ref{S16a}b). However, at the lowest temperatures $I_c$ appears to be temperature-independent. The magnetoresistance (MR) of sample C exhibits typical features of field induced SC-to-normal state transition (Figs.~\ref{S16b}a and \ref{S16b}b). As the temperature increases, the zero-field resistance becomes higher, although the shape of MR curves remains similar. We clearly observe the anisotropy of critical magnetic field -- $H_c$ values for longitudinal field are considerably higher than those for transverse field, which is common for thin SC films.

Assuming that $H_c$ corresponds to the 50$\%$ resistance drop of its normal value (midpoint), we obtained $H_c-T_c$ diagrams for the studied samples (Figs.~\ref{S38}b, \ref{S58}b{\red ,} and \ref{S16b}c). To characterize the observed SC state, we applied the conventional formula describing the decrease in the critical magnetic field, $H_c$, upon increasing the temperature below critical value $T_c$:
\begin{equation}
\label{crit}
H(T)=H_c(0)\cdot \left ( 1- \left ( \frac{T}{T_c}\right )^\alpha \right ) ^\beta.
\end{equation}
Here, $\alpha=2;\beta=1$ corresponds to the common Bardeen-Cooper-Schrieffer (BCS) theory at low temperatures, while the Ginzburg-Landau (G-L) theory suggests $\alpha=1;\beta=1$ to describe the temperature dependence of upper critical field, $H_{c2}$, in type-II SC close to $T_c$. It is important to note that the presented $H_c-T_c$ diagrams for studied samples cannot be approximated well by Eq. (\ref{crit}) with $\alpha=2;\beta=1$. However, in the intermediate temperature range, the obtained diagrams can be effectively described with linear function ($\alpha=1;\beta=1$) as it is shown in the corresponding figures. The deviations from this line in the vicinity of $T_c$ (samples A and B) and low-temperature saturation (sample C) requires additional investigation and can be related to the polycrystallinity of studied films. Nevertheless, the overall behavior of $H_c-T_c$ diagrams for transverse and longitudinal magnetic fields is essentially the same. It is worth mentioning that there is no qualitative difference if we ascribe $T_c$ and $H_c$ to the 90$\%$ value of normal state resistance (see Fig.~\ref{S38}b).

The absence of elemental segregation (confirmed by EDXS results) implies that the observed SC phase cannot be attributed to the elemental Cd, which is a classical SC with $T_c=0.56$ K and $H_c\approx 30$ Oe~\cite{Clem1953,good1951} (also, the $H_c$ values for studied films are substantially higher, e.g. for sample A the transverse $H_c\approx350$ Oe). Additionally, due to relatively large sizes of conducting channel, the observed SC cannot be related to an influence of the contact regions. Thus, we can conclude that the observed SC phase emerges within Cd-As binary system with the elemental ratio close to stoichiometric Cd$_3$As$_2$ compound. Alongside with the observed features in Raman spectra (inherent to the Cd$_3$As$_2$ crystal phase), this suggests a possible coexistence of SC and DSM phase within studied films, which can support the formation of surface Majoranna modes~\cite{SatoPRL2015,SatoPRB2016}.

Originally, the G-L theory describes SC in the vicinity of $T_c$. However, fairly often it is applied in the much broader temperature range~\cite{Yan2018}. In our study, application of the G-L theory implies that transverse $H_c$ values extrapolated to zero temperature can be used for the estimation of coherence length $\xi$ ($H_c(0)=\phi_0/(2\pi \xi ^2)$, $\phi_0$ - magnetic flux quantum). For sample C, we obtain $\xi\approx 100$ nm (for the midpoint $H_c$ value), which is twice as high as the film thickness. The latter suggests that the observed SC should be two-dimensional, and the corresponding $H_c-T_c$ diagram for longitudinal field should follow Eq.~(\ref{crit}) with $\alpha=1;\beta=1/2$~\cite{chopra1979}. The temperature range, where this type of $H_c-T_c$ correlation should be observed, depends on various parameters of the film~\cite{Shapoval1966JETP}. Thus, the absence of such behavior for studied films suggest that it may be very narrow. On the other hand, the isotropic linear character of $H_c(T)$ dependeces may also indicate that some additional effects are relevant in our case. A more general approach, described in Ref.~\cite{Kuch2017JETP}, suggests that linear $H_c-T_c$ relation in the wide temperature range can be observed for systems with strong pairing interaction. According to the phase diagram of SC in Cd$_3$As$_2$~\cite{SatoPRB2016,SatoRPP2017}, the type of pairing potential (even- or odd-parity) depends not on the integral strength of \emph{e--e} attraction, but rather on the relation between intra- and inter-orbital components. However, in the studies of pressure-induced SC in Cd$_3$As$_2$~\cite{HeNPJQuMat2016} and Bi$_2$Se$_3$~\cite{kirsh2013,kong2013}, the linear character of $H_c-T_c$ diagram (alongside with the saturation of pressure dependence of $T_c$) was considered as a signature of nontrivial pairing (of odd-parity). Thus, the obtained $H_c-T_c$ diagrams with pronounced linear behavior may indicate the presence of odd-parity pairing potential in studied films, which, of course, requires more detailed experimental verification.

From the conventional point of view, the SC phase emergence in our case is favored due to high electron densities (the Hall concentrations for studied films are about $8\div12\cdot10^{18}$ cm$^{-3}$). It implies that the Fermi level lies substantially higher than the Lifshitz transition of pristine Cd$_3$As$_2$ single crystals, meaning that transport features of DSM phase might be damped, although the band structure may preserve the nonzero Berry curvature~\cite{Nish2018PRB}. Nevertheless, the substantial increase in the electron density in topological materials always precedes the SC phase emergence, which, however, is argued to sustain topological features~\cite{HeNPJQuMat2016,kirsh2013,kong2013}. Considering the fact that high quality Cd$_3$As$_2$ single crystalline films studied in similar temperature range did not exhibit any SC transition~\cite{gall2018}, we assume that the SC phase in our case results from possible crystal structure distortions.

Knowing that intrinsic SC emergence in Cd$_3$As$_2$ is induced by pressure~\cite{HeNPJQuMat2016,WangNatMat2016,AggarwalNatMat2016}, we also tend to assume that the observed SC can be affected by strain existing within a film (e.g. due to polycrystalline character). Due to relatively small thickness of the studied films, such strain can be also related to the difference of the coefficients of thermal expansion (CTE) of the substrate and the film. At room temperature, Cd$_3$As$_2$ is characterized by large CTE value of about $\alpha_{CdAs} \sim 11.8 \div 12.4 \cdot 10^{-6}\,$K$^{-1}$. It decreases with the temperature as it is for normal metal~\cite{Pietraszko1973}. Al$_2$O$_3$ has anisotropic CTE (from $\alpha_\parallel = 6.7\cdot 10^{-6}\,$K$^{-1}$ to $\alpha_\perp = 5.0 \cdot 10^{-6}\,$K$^{-1}$) at room temperature~\cite{Mejlikhov1991}. Si has smaller CTE at room temperatures ($\alpha_{Si} = 2.54 \cdot 10^{-6}\,$K$^{-1}$) and exhibits even negative values at lower temperatures, $110 \div 20\,$K~\cite{Novikova1974}. Thus, as the temperature decreases, the Al$_2$O$_3$ substrate contracts less, and the Si substrate contracts even much less than the Cd$_3$As$_2$ film. As a result, studied samples should experience the tensile strain in the temperature range under study. We did not manage to find either the experimental or theoretical studies of Cd$_3$As$_2$ phase diagram under negative pressures, making it hard to predict the effect of such strain. We observed close values of $T_c\approx 190$ mK (midpoint) for samples A and B with substantially different width of SC transition. Moreover, preliminary studies of films deposited at the same conditions on the fused quartz substrates (with $\alpha \ll \alpha_{Si}$) revealed the absence of SC transition down to 40 mK. Thus, it is clear that the substrate strongly affects the emergent SC state in studied films. However, to elucidate all relevant effects, an additional investigation is highly needed.

It is important to note that the indications of surface SC in Cd$_3$As$_2$ single crystals was recently reported~\cite{Shvetsov2018arxive}. The apparent difference is that we observed almost zero resistance of the whole film, while the authors of Ref.~\cite{Shvetsov2018arxive} dealt with the 10$\%$ drop of differential resistance measured at the sample surface. Whereas this difference is probably related to the small thickness of studied films, the properties of Majorana modes in our case can be different from those of single DSM surface~\cite{SatoPRB2016} due to Weyl orbit formation.

In this Letter, we report the first experimental observation of superconductivity emergence in cadmium arsenide films without any applied pressure. The superconducting nature of the transition observed in $R(T)$ dependence is justified by differential resistance and magnetoresistance measurements, providing temperature dependence of corresponding critical parameters ($I_c$ and $H_c$). We argue that the observed phenomena cannot be attributed to any parasitic effects such as elemental Cd segregation or any effects related to contact regions. The observed SC state is characterized by $H_c-T_c$ diagrams with a pronounced linear regions at intermediate temperatures for both the transverse and longitudinal magnetic fields. The deviations from overall linear dependence at low temperatures and close to $T_c$ can be related to the polycrystallinity of studied films. Similar linear $H_c(T)$ dependence was considered as a signature of nontrivial pairing potential in bulk Cd$_3$As$_2$ and Bi$_2$Se$_3$ films under pressure. Theory suggests that such pairing potential should result in the formation of surface Majoranna modes. Favored by high electron densities, the observed superconducting state can emerge due to distortion of the film crystal structure, which arises during the deposition procedure using the magnetron sputtering. We argue that this SC state can be affected by various strains arising in the films under study. We also note that similar films deposited on fused quartz substrates do not exhibit the superconducting state emergence down to 40 mK. Thus, it is possible to substantially affect the observed SC state. The latter indicates that the investigated systems might be a promising platform for studies of topological superconductivity.

\subsection*{Acknowledgements}

We thank K.V. Mitsen and O.M. Ivanenko for fruitful discussions. The work was partially supported by the Russian Science Foundation, grant No. 17-12-01345. A portion of this work was performed at the National High Magnetic Field Laboratory, which is supported by the National Science Foundation Cooperative Agreement No. DMR-1644779 and the State of Florida. A portion of theoretical analysis, performed by V.M.P., was supported by Russian Foundation for Basic Research, grant No. 16-29-03330.

\bibliographystyle{apsrevlong_no_issn_url}
\bibliography{FilmsBIB}

\begin{thebibliography}{41}
\expandafter\ifx\csname natexlab\endcsname\relax\def\natexlab#1{#1}\fi
\expandafter\ifx\csname bibnamefont\endcsname\relax
  \def\bibnamefont#1{#1}\fi
\expandafter\ifx\csname bibfnamefont\endcsname\relax
  \def\bibfnamefont#1{#1}\fi
\expandafter\ifx\csname citenamefont\endcsname\relax
  \def\citenamefont#1{#1}\fi

\bibitem[{\citenamefont{Armitage et~al.}(2018)\citenamefont{Armitage, Mele, and
  Vishwanath}}]{ArmitageRMP2018}
\bibinfo{author}{\bibfnamefont{N.~P.} \bibnamefont{Armitage}},
  \bibinfo{author}{\bibfnamefont{E.~J.} \bibnamefont{Mele}}, \bibnamefont{and}
  \bibinfo{author}{\bibfnamefont{A.}~\bibnamefont{Vishwanath}},
  {``}\bibinfo{title}{Weyl and Dirac semimetals in three-dimensional
  solids},{''} \bibinfo{journal}{Rev. Mod. Phys.}
  \textbf{\bibinfo{volume}{90}}, \bibinfo{pages}{015001}
  (\bibinfo{year}{2018}).

\bibitem[{\citenamefont{Yan and Felser}(2017)}]{YanAnnRevCMPh2017}
\bibinfo{author}{\bibfnamefont{B.}~\bibnamefont{Yan}} \bibnamefont{and}
  \bibinfo{author}{\bibfnamefont{C.}~\bibnamefont{Felser}},
  {``}\bibinfo{title}{Topological materials: {Weyl} semimetals},{''}
  \bibinfo{journal}{Annu. Rev. Condens. Matter Phys.}
  \textbf{\bibinfo{volume}{8}}, \bibinfo{pages}{337} (\bibinfo{year}{2017}).

\bibitem[{\citenamefont{Wang et~al.}(2017)\citenamefont{Wang, Lin, Wang, Yu,
  and Liao}}]{WangAdvPhX2017}
\bibinfo{author}{\bibfnamefont{S.}~\bibnamefont{Wang}},
  \bibinfo{author}{\bibfnamefont{B.-C.} \bibnamefont{Lin}},
  \bibinfo{author}{\bibfnamefont{A.-Q.} \bibnamefont{Wang}},
  \bibinfo{author}{\bibfnamefont{D.-P.} \bibnamefont{Yu}}, \bibnamefont{and}
  \bibinfo{author}{\bibfnamefont{Z.-M.} \bibnamefont{Liao}},
  {``}\bibinfo{title}{Quantum transport in Dirac and Weyl semimetals: a
  review},{''} \bibinfo{journal}{Adv. Phys. X} \textbf{\bibinfo{volume}{2}},
  \bibinfo{pages}{518} (\bibinfo{year}{2017}).

\bibitem[{\citenamefont{Neupane et~al.}(2013)\citenamefont{Neupane, Xu, Sankar,
  Alidoust, Bian, Liu, Belopolski, Chang, Jeng, Lin
  et~al.}}]{Neupane_NatCom2013}
\bibinfo{author}{\bibfnamefont{M.}~\bibnamefont{Neupane}},
  \bibinfo{author}{\bibfnamefont{S.~Y.} \bibnamefont{Xu}},
  \bibinfo{author}{\bibfnamefont{R.}~\bibnamefont{Sankar}},
  \bibinfo{author}{\bibfnamefont{N.}~\bibnamefont{Alidoust}},
  \bibinfo{author}{\bibfnamefont{G.}~\bibnamefont{Bian}},
  \bibinfo{author}{\bibfnamefont{C.}~\bibnamefont{Liu}},
  \bibinfo{author}{\bibfnamefont{I.}~\bibnamefont{Belopolski}},
  \bibinfo{author}{\bibfnamefont{T.~R.} \bibnamefont{Chang}},
  \bibinfo{author}{\bibfnamefont{H.~T.} \bibnamefont{Jeng}},
  \bibinfo{author}{\bibfnamefont{H.}~\bibnamefont{Lin}}, \bibnamefont{et~al.},
  {``}\bibinfo{title}{Observation of a three-dimensional topological Dirac
  semimetal phase in high-mobility Cd$_3$As$_2$},{''} \bibinfo{journal}{Nat.
  Commun.} \textbf{\bibinfo{volume}{5}}, \bibinfo{pages}{3786}
  (\bibinfo{year}{2013}).

\bibitem[{\citenamefont{Liu et~al.}(2014)\citenamefont{Liu, Jiang, Zhou, Wang,
  Zhang, Weng, Prabhakaran, Mo, Peng, Dudin et~al.}}]{Liu_NatMat2014}
\bibinfo{author}{\bibfnamefont{Z.~K.} \bibnamefont{Liu}},
  \bibinfo{author}{\bibfnamefont{J.}~\bibnamefont{Jiang}},
  \bibinfo{author}{\bibfnamefont{B.}~\bibnamefont{Zhou}},
  \bibinfo{author}{\bibfnamefont{Z.~J.} \bibnamefont{Wang}},
  \bibinfo{author}{\bibfnamefont{Y.}~\bibnamefont{Zhang}},
  \bibinfo{author}{\bibfnamefont{H.~M.} \bibnamefont{Weng}},
  \bibinfo{author}{\bibfnamefont{D.}~\bibnamefont{Prabhakaran}},
  \bibinfo{author}{\bibfnamefont{S.~K.} \bibnamefont{Mo}},
  \bibinfo{author}{\bibfnamefont{H.}~\bibnamefont{Peng}},
  \bibinfo{author}{\bibfnamefont{P.}~\bibnamefont{Dudin}},
  \bibnamefont{et~al.}, {``}\bibinfo{title}{A stable three-dimensional
  topological Dirac semimetal Cd$_3$As$_2$},{''} \bibinfo{journal}{Nat. Mater.}
  \textbf{\bibinfo{volume}{13}}, \bibinfo{pages}{677} (\bibinfo{year}{2014}).

\bibitem[{\citenamefont{Jeon et~al.}(2014)\citenamefont{Jeon, Zhou, Gyenis,
  Feldman, Kimchi, Potter, Gibson, Cava, Vishwanath, and
  Yazdani}}]{Jeon_NatMat2014}
\bibinfo{author}{\bibfnamefont{S.}~\bibnamefont{Jeon}},
  \bibinfo{author}{\bibfnamefont{B.~B.} \bibnamefont{Zhou}},
  \bibinfo{author}{\bibfnamefont{A.}~\bibnamefont{Gyenis}},
  \bibinfo{author}{\bibfnamefont{B.~E.} \bibnamefont{Feldman}},
  \bibinfo{author}{\bibfnamefont{I.}~\bibnamefont{Kimchi}},
  \bibinfo{author}{\bibfnamefont{A.~C.} \bibnamefont{Potter}},
  \bibinfo{author}{\bibfnamefont{Q.~D.} \bibnamefont{Gibson}},
  \bibinfo{author}{\bibfnamefont{R.~J.} \bibnamefont{Cava}},
  \bibinfo{author}{\bibfnamefont{A.}~\bibnamefont{Vishwanath}},
  \bibnamefont{and} \bibinfo{author}{\bibfnamefont{A.}~\bibnamefont{Yazdani}},
  {``}\bibinfo{title}{Landau quantization and quasiparticle interference in the
  three-dimensional Dirac semimetal Cd$_3$As$_2$},{''} \bibinfo{journal}{Nat.
  Mater.} \textbf{\bibinfo{volume}{13}}, \bibinfo{pages}{677}
  (\bibinfo{year}{2014}).

\bibitem[{\citenamefont{Borisenko et~al.}(2014)\citenamefont{Borisenko, Gibson,
  Evtushinsky, Zabolotnyy, B\"uchner, and Cava}}]{Borisenko_PRL2014}
\bibinfo{author}{\bibfnamefont{S.}~\bibnamefont{Borisenko}},
  \bibinfo{author}{\bibfnamefont{Q.}~\bibnamefont{Gibson}},
  \bibinfo{author}{\bibfnamefont{D.}~\bibnamefont{Evtushinsky}},
  \bibinfo{author}{\bibfnamefont{V.}~\bibnamefont{Zabolotnyy}},
  \bibinfo{author}{\bibfnamefont{B.}~\bibnamefont{B\"uchner}},
  \bibnamefont{and} \bibinfo{author}{\bibfnamefont{R.~J.} \bibnamefont{Cava}},
  {``}\bibinfo{title}{Experimental Realization of a Three-Dimensional Dirac
  Semimetal},{''} \bibinfo{journal}{Phys. Rev. Lett.}
  \textbf{\bibinfo{volume}{113}}, \bibinfo{pages}{027603}
  (\bibinfo{year}{2014}).

\bibitem[{\citenamefont{Lu et~al.}(2015)\citenamefont{Lu, Wang, Ye, Ran, Fu,
  Joannopoulos, and Solja{\v c}i{\'c}}}]{LuScience2015}
\bibinfo{author}{\bibfnamefont{L.}~\bibnamefont{Lu}},
  \bibinfo{author}{\bibfnamefont{Z.}~\bibnamefont{Wang}},
  \bibinfo{author}{\bibfnamefont{D.}~\bibnamefont{Ye}},
  \bibinfo{author}{\bibfnamefont{L.}~\bibnamefont{Ran}},
  \bibinfo{author}{\bibfnamefont{L.}~\bibnamefont{Fu}},
  \bibinfo{author}{\bibfnamefont{J.~D.} \bibnamefont{Joannopoulos}},
  \bibnamefont{and} \bibinfo{author}{\bibfnamefont{M.}~\bibnamefont{Solja{\v
  c}i{\'c}}}, {``}\bibinfo{title}{Experimental observation of Weyl points},{''}
  \bibinfo{journal}{Science} \textbf{\bibinfo{volume}{349}},
  \bibinfo{pages}{622} (\bibinfo{year}{2015}).

\bibitem[{\citenamefont{Dubowski and Williams}(1984)}]{DubowskiAPL1984}
\bibinfo{author}{\bibfnamefont{J.~J.} \bibnamefont{Dubowski}} \bibnamefont{and}
  \bibinfo{author}{\bibfnamefont{D.~F.} \bibnamefont{Williams}},
  {``}\bibinfo{title}{Pulsed laser evaporation of Cd$_3$As$_2$},{''}
  \bibinfo{journal}{Appl. Phys. Lett.} \textbf{\bibinfo{volume}{44}},
  \bibinfo{pages}{339} (\bibinfo{year}{1984}).

\bibitem[{\citenamefont{Weclewicz and Zdanowicz}(1987)}]{WeclewiczTSF1987}
\bibinfo{author}{\bibfnamefont{C.}~\bibnamefont{Weclewicz}} \bibnamefont{and}
  \bibinfo{author}{\bibfnamefont{L.}~\bibnamefont{Zdanowicz}},
  {``}\bibinfo{title}{Transport properties of thin amorphous films of cadmium
  arsenide},{''} \bibinfo{journal}{Thin Solid Films}
  \textbf{\bibinfo{volume}{151}}, \bibinfo{pages}{87 } (\bibinfo{year}{1987}).

\bibitem[{\citenamefont{El-Shazly et~al.}(1996)\citenamefont{El-Shazly,
  Soliman, Abd El-Hady, and El-Sayed}}]{ShazlyVac1996}
\bibinfo{author}{\bibfnamefont{A.~A.} \bibnamefont{El-Shazly}},
  \bibinfo{author}{\bibfnamefont{H.}~\bibnamefont{Soliman}},
  \bibinfo{author}{\bibfnamefont{D.}~\bibnamefont{Abd El-Hady}},
  \bibnamefont{and} \bibinfo{author}{\bibfnamefont{H.~E.~A.}
  \bibnamefont{El-Sayed}}, {``}\bibinfo{title}{Transport properties of thin
  Cd$_3$As$_2$ polycrystalline films},{''} \bibinfo{journal}{Vacuum}
  \textbf{\bibinfo{volume}{47}}, \bibinfo{pages}{45 } (\bibinfo{year}{1996}).

\bibitem[{\citenamefont{Jarzabek et~al.}(2004)\citenamefont{Jarzabek, Weszka, ,
  and Cisowski}}]{JarzabekJNCS2004}
\bibinfo{author}{\bibfnamefont{B.}~\bibnamefont{Jarzabek}},
  \bibinfo{author}{\bibfnamefont{J.}~\bibnamefont{Weszka}}, , \bibnamefont{and}
  \bibinfo{author}{\bibfnamefont{J.}~\bibnamefont{Cisowski}},
  {``}\bibinfo{title}{Distribution of electronic states in amorphous Cd-As thin
  films on the basis of optical measurements},{''} \bibinfo{journal}{J.
  Non-Cryst. Solids} \textbf{\bibinfo{volume}{333}}, \bibinfo{pages}{206 }
  (\bibinfo{year}{2004}).

\bibitem[{\citenamefont{Wray et~al.}(2010)\citenamefont{Wray, Xu, Xia, Hor,
  Qian, Fedorov, Lin, Bansil, Cava, and Hasan}}]{WrayNatPh2010}
\bibinfo{author}{\bibfnamefont{L.~A.} \bibnamefont{Wray}},
  \bibinfo{author}{\bibfnamefont{S.-Y.} \bibnamefont{Xu}},
  \bibinfo{author}{\bibfnamefont{Y.}~\bibnamefont{Xia}},
  \bibinfo{author}{\bibfnamefont{Y.~S.} \bibnamefont{Hor}},
  \bibinfo{author}{\bibfnamefont{D.}~\bibnamefont{Qian}},
  \bibinfo{author}{\bibfnamefont{A.~V.} \bibnamefont{Fedorov}},
  \bibinfo{author}{\bibfnamefont{H.}~\bibnamefont{Lin}},
  \bibinfo{author}{\bibfnamefont{A.}~\bibnamefont{Bansil}},
  \bibinfo{author}{\bibfnamefont{R.~J.} \bibnamefont{Cava}}, \bibnamefont{and}
  \bibinfo{author}{\bibfnamefont{M.~Z.} \bibnamefont{Hasan}},
  {``}\bibinfo{title}{Observation of topological order in a superconducting
  doped topological insulator},{''} \bibinfo{journal}{Nat. Phys.}
  \textbf{\bibinfo{volume}{6}}, \bibinfo{pages}{855} (\bibinfo{year}{2010}).

\bibitem[{\citenamefont{Ando and Fu}(2015)}]{AndoAnnRevCMPh2015}
\bibinfo{author}{\bibfnamefont{Y.}~\bibnamefont{Ando}} \bibnamefont{and}
  \bibinfo{author}{\bibfnamefont{L.}~\bibnamefont{Fu}},
  {``}\bibinfo{title}{Topological crystalline insulators and topological
  superconductors: {From} concepts to materials},{''} \bibinfo{journal}{Annu.l
  Rev. Condens. Matter Phys.} \textbf{\bibinfo{volume}{6}},
  \bibinfo{pages}{361} (\bibinfo{year}{2015}).

\bibitem[{\citenamefont{Sato and Ando}(2017)}]{SatoRPP2017}
\bibinfo{author}{\bibfnamefont{M.}~\bibnamefont{Sato}} \bibnamefont{and}
  \bibinfo{author}{\bibfnamefont{Y.}~\bibnamefont{Ando}},
  {``}\bibinfo{title}{Topological superconductors: a review},{''}
  \bibinfo{journal}{Rep. Prog. Phys.} \textbf{\bibinfo{volume}{80}},
  \bibinfo{pages}{076501} (\bibinfo{year}{2017}).

\bibitem[{\citenamefont{Kobayashi and Sato}(2015)}]{SatoPRL2015}
\bibinfo{author}{\bibfnamefont{S.}~\bibnamefont{Kobayashi}} \bibnamefont{and}
  \bibinfo{author}{\bibfnamefont{M.}~\bibnamefont{Sato}},
  {``}\bibinfo{title}{Topological Superconductivity in Dirac Semimetals},{''}
  \bibinfo{journal}{Phys. Rev. Lett.} \textbf{\bibinfo{volume}{115}},
  \bibinfo{pages}{187001} (\bibinfo{year}{2015}).

\bibitem[{\citenamefont{Hashimoto et~al.}(2016)\citenamefont{Hashimoto,
  Kobayashi, Tanaka, and Sato}}]{SatoPRB2016}
\bibinfo{author}{\bibfnamefont{T.}~\bibnamefont{Hashimoto}},
  \bibinfo{author}{\bibfnamefont{S.}~\bibnamefont{Kobayashi}},
  \bibinfo{author}{\bibfnamefont{Y.}~\bibnamefont{Tanaka}}, \bibnamefont{and}
  \bibinfo{author}{\bibfnamefont{M.}~\bibnamefont{Sato}},
  {``}\bibinfo{title}{Superconductivity in doped Dirac semimetals},{''}
  \bibinfo{journal}{Phys. Rev. B} \textbf{\bibinfo{volume}{94}},
  \bibinfo{pages}{014510} (\bibinfo{year}{2016}).

\bibitem[{\citenamefont{He et~al.}(2016)\citenamefont{He, Jia, Zhang, Hong,
  Jin, and Li}}]{HeNPJQuMat2016}
\bibinfo{author}{\bibfnamefont{L.}~\bibnamefont{He}},
  \bibinfo{author}{\bibfnamefont{Y.}~\bibnamefont{Jia}},
  \bibinfo{author}{\bibfnamefont{S.}~\bibnamefont{Zhang}},
  \bibinfo{author}{\bibfnamefont{X.}~\bibnamefont{Hong}},
  \bibinfo{author}{\bibfnamefont{C.}~\bibnamefont{Jin}}, \bibnamefont{and}
  \bibinfo{author}{\bibfnamefont{S.}~\bibnamefont{Li}},
  {``}\bibinfo{title}{Pressure-induced superconductivity in the
  three-dimensional topological Dirac semimetal Cd$_3$As$_2$},{''}
  \bibinfo{journal}{Npj Quant. Mater.} \textbf{\bibinfo{volume}{1}},
  \bibinfo{pages}{16014} (\bibinfo{year}{2016}).

\bibitem[{\citenamefont{Wang et~al.}(2015)\citenamefont{Wang, Wang, Liu, Lu,
  Yang, Jia, Liu, Xie, Wei, and Wang}}]{WangNatMat2016}
\bibinfo{author}{\bibfnamefont{H.}~\bibnamefont{Wang}},
  \bibinfo{author}{\bibfnamefont{H.}~\bibnamefont{Wang}},
  \bibinfo{author}{\bibfnamefont{H.}~\bibnamefont{Liu}},
  \bibinfo{author}{\bibfnamefont{H.}~\bibnamefont{Lu}},
  \bibinfo{author}{\bibfnamefont{W.}~\bibnamefont{Yang}},
  \bibinfo{author}{\bibfnamefont{S.}~\bibnamefont{Jia}},
  \bibinfo{author}{\bibfnamefont{X.-J.} \bibnamefont{Liu}},
  \bibinfo{author}{\bibfnamefont{X.~C.} \bibnamefont{Xie}},
  \bibinfo{author}{\bibfnamefont{J.}~\bibnamefont{Wei}}, \bibnamefont{and}
  \bibinfo{author}{\bibfnamefont{J.}~\bibnamefont{Wang}},
  {``}\bibinfo{title}{Observation of superconductivity induced by a point
  contact on 3D Dirac semimetal Cd$_3$As$_2$ crystals},{''}
  \bibinfo{journal}{Nat. Mater.} \textbf{\bibinfo{volume}{15}},
  \bibinfo{pages}{38} (\bibinfo{year}{2015}).

\bibitem[{\citenamefont{Aggarwal et~al.}(2015)\citenamefont{Aggarwal, Gaurav,
  Thakur, Haque, Ganguli, and Sheet}}]{AggarwalNatMat2016}
\bibinfo{author}{\bibfnamefont{L.}~\bibnamefont{Aggarwal}},
  \bibinfo{author}{\bibfnamefont{A.}~\bibnamefont{Gaurav}},
  \bibinfo{author}{\bibfnamefont{G.~S.} \bibnamefont{Thakur}},
  \bibinfo{author}{\bibfnamefont{Z.}~\bibnamefont{Haque}},
  \bibinfo{author}{\bibfnamefont{A.~K.} \bibnamefont{Ganguli}},
  \bibnamefont{and} \bibinfo{author}{\bibfnamefont{G.}~\bibnamefont{Sheet}},
  {``}\bibinfo{title}{Unconventional superconductivity at mesoscopic point
  contacts on the 3D Dirac semimetal Cd$_3$As$_2$},{''} \bibinfo{journal}{Nat.
  Mater.} \textbf{\bibinfo{volume}{15}}, \bibinfo{pages}{32}
  (\bibinfo{year}{2015}).

\bibitem[{\citenamefont{Li et~al.}(2018)\citenamefont{Li, Li, Wang, Wang, Liao,
  Brinkman, and Yu}}]{LiPRB2018}
\bibinfo{author}{\bibfnamefont{C.-Z.} \bibnamefont{Li}},
  \bibinfo{author}{\bibfnamefont{C.}~\bibnamefont{Li}},
  \bibinfo{author}{\bibfnamefont{L.-X.} \bibnamefont{Wang}},
  \bibinfo{author}{\bibfnamefont{S.}~\bibnamefont{Wang}},
  \bibinfo{author}{\bibfnamefont{Z.-M.} \bibnamefont{Liao}},
  \bibinfo{author}{\bibfnamefont{A.}~\bibnamefont{Brinkman}}, \bibnamefont{and}
  \bibinfo{author}{\bibfnamefont{D.-P.} \bibnamefont{Yu}},
  {``}\bibinfo{title}{Bulk and surface states carried supercurrent in ballistic
  Nb-Dirac semimetal ${\mathrm{Cd}}_{3}{\mathrm{As}}_{2}$ nanowire-Nb
  junctions},{''} \bibinfo{journal}{Phys. Rev. B}
  \textbf{\bibinfo{volume}{97}}, \bibinfo{pages}{115446}
  (\bibinfo{year}{2018}).

\bibitem[{\citenamefont{Yu et~al.}(2018)\citenamefont{Yu, Pan, Medlin,
  Rodriguez, Lee, Bao, and Zhang}}]{YuPRL2018}
\bibinfo{author}{\bibfnamefont{W.}~\bibnamefont{Yu}},
  \bibinfo{author}{\bibfnamefont{W.}~\bibnamefont{Pan}},
  \bibinfo{author}{\bibfnamefont{D.~L.} \bibnamefont{Medlin}},
  \bibinfo{author}{\bibfnamefont{M.~A.} \bibnamefont{Rodriguez}},
  \bibinfo{author}{\bibfnamefont{S.~R.} \bibnamefont{Lee}},
  \bibinfo{author}{\bibfnamefont{Z.-q.} \bibnamefont{Bao}}, \bibnamefont{and}
  \bibinfo{author}{\bibfnamefont{F.}~\bibnamefont{Zhang}},
  {``}\bibinfo{title}{$\ensuremath{\pi}$ and $4\ensuremath{\pi}$ Josephson
  Effects Mediated by a Dirac Semimetal},{''} \bibinfo{journal}{Phys. Rev.
  Lett.} \textbf{\bibinfo{volume}{120}}, \bibinfo{pages}{177704}
  (\bibinfo{year}{2018}).

\bibitem[{\citenamefont{Uchida et~al.}(2017)\citenamefont{Uchida, Nakazawa,
  Nishihaya, Akiba, Kriener, Kozuka, Miyake, Taguchi, Tokunaga, Nagaosa
  et~al.}}]{Uchida2017Natcomm}
\bibinfo{author}{\bibfnamefont{M.}~\bibnamefont{Uchida}},
  \bibinfo{author}{\bibfnamefont{Y.}~\bibnamefont{Nakazawa}},
  \bibinfo{author}{\bibfnamefont{S.}~\bibnamefont{Nishihaya}},
  \bibinfo{author}{\bibfnamefont{K.}~\bibnamefont{Akiba}},
  \bibinfo{author}{\bibfnamefont{M.}~\bibnamefont{Kriener}},
  \bibinfo{author}{\bibfnamefont{Y.}~\bibnamefont{Kozuka}},
  \bibinfo{author}{\bibfnamefont{A.}~\bibnamefont{Miyake}},
  \bibinfo{author}{\bibfnamefont{Y.}~\bibnamefont{Taguchi}},
  \bibinfo{author}{\bibfnamefont{M.}~\bibnamefont{Tokunaga}},
  \bibinfo{author}{\bibfnamefont{N.}~\bibnamefont{Nagaosa}},
  \bibnamefont{et~al.}, {``}\bibinfo{title}{Quantum Hall states observed in
  thin films of Dirac semimetal ${\mathrm{Cd}}_{3}{\mathrm{As}}_{2}$},{''}
  \bibinfo{journal}{Nat. Commun.} \textbf{\bibinfo{volume}{8}},
  \bibinfo{pages}{2274} (\bibinfo{year}{2017}).

\bibitem[{\citenamefont{Schumann et~al.}(2018)\citenamefont{Schumann, Galletti,
  Kealhofer, Kim, Goyal, and Stemmer}}]{Schumann2018PRL}
\bibinfo{author}{\bibfnamefont{T.}~\bibnamefont{Schumann}},
  \bibinfo{author}{\bibfnamefont{L.}~\bibnamefont{Galletti}},
  \bibinfo{author}{\bibfnamefont{D.~A.} \bibnamefont{Kealhofer}},
  \bibinfo{author}{\bibfnamefont{H.}~\bibnamefont{Kim}},
  \bibinfo{author}{\bibfnamefont{M.}~\bibnamefont{Goyal}}, \bibnamefont{and}
  \bibinfo{author}{\bibfnamefont{S.}~\bibnamefont{Stemmer}},
  {``}\bibinfo{title}{Observation of the quantum Hall effect in confined films
  of the three-dimensional Dirac semimetal
  ${\mathrm{Cd}}_{3}{\mathrm{As}}_{2}$},{''} \bibinfo{journal}{Phys. Rev.
  Lett.} \textbf{\bibinfo{volume}{120}}, \bibinfo{pages}{016801}
  (\bibinfo{year}{2018}).

\bibitem[{\citenamefont{Zhang et~al.}(2017)\citenamefont{Zhang, Narayan, Lu,
  Zhang, Zhang, Ni, Yuan, Liu, Park, Zhang et~al.}}]{Zhang2017NatComm}
\bibinfo{author}{\bibfnamefont{C.}~\bibnamefont{Zhang}},
  \bibinfo{author}{\bibfnamefont{A.}~\bibnamefont{Narayan}},
  \bibinfo{author}{\bibfnamefont{S.}~\bibnamefont{Lu}},
  \bibinfo{author}{\bibfnamefont{J.}~\bibnamefont{Zhang}},
  \bibinfo{author}{\bibfnamefont{H.}~\bibnamefont{Zhang}},
  \bibinfo{author}{\bibfnamefont{Z.}~\bibnamefont{Ni}},
  \bibinfo{author}{\bibfnamefont{X.}~\bibnamefont{Yuan}},
  \bibinfo{author}{\bibfnamefont{Y.}~\bibnamefont{Liu}},
  \bibinfo{author}{\bibfnamefont{J.-H.} \bibnamefont{Park}},
  \bibinfo{author}{\bibfnamefont{E.}~\bibnamefont{Zhang}},
  \bibnamefont{et~al.}, {``}\bibinfo{title}{Evolution of Weyl orbit and quantum
  Hall effect in Dirac semimetal ${\mathrm{Cd}}_{3}{\mathrm{As}}_{2}$},{''}
  \bibinfo{journal}{Nat. Commun.} \textbf{\bibinfo{volume}{8}},
  \bibinfo{pages}{1272} (\bibinfo{year}{2017}).

\bibitem[{\citenamefont{Galletti et~al.}(2018)\citenamefont{Galletti, Schumann,
  Shoron, Goyal, Kealhofer, Kim, and Stemmer}}]{gall2018}
\bibinfo{author}{\bibfnamefont{L.}~\bibnamefont{Galletti}},
  \bibinfo{author}{\bibfnamefont{T.}~\bibnamefont{Schumann}},
  \bibinfo{author}{\bibfnamefont{O.~F.} \bibnamefont{Shoron}},
  \bibinfo{author}{\bibfnamefont{M.}~\bibnamefont{Goyal}},
  \bibinfo{author}{\bibfnamefont{D.~A.} \bibnamefont{Kealhofer}},
  \bibinfo{author}{\bibfnamefont{H.}~\bibnamefont{Kim}}, \bibnamefont{and}
  \bibinfo{author}{\bibfnamefont{S.}~\bibnamefont{Stemmer}},
  {``}\bibinfo{title}{Two-dimensional Dirac fermions in thin films of
  $\mathrm{C}{\mathrm{d}}_{3}\mathrm{A}{\mathrm{s}}_{2}$},{''}
  \bibinfo{journal}{Phys. Rev. B} \textbf{\bibinfo{volume}{97}},
  \bibinfo{pages}{115132} (\bibinfo{year}{2018}).

\bibitem[{\citenamefont{Wei et~al.}(2006)\citenamefont{Wei, Lu, Yu, Zhang, and
  Qian}}]{Wei2006CGD}
\bibinfo{author}{\bibfnamefont{S.}~\bibnamefont{Wei}},
  \bibinfo{author}{\bibfnamefont{J.}~\bibnamefont{Lu}},
  \bibinfo{author}{\bibfnamefont{W.}~\bibnamefont{Yu}},
  \bibinfo{author}{\bibfnamefont{H.}~\bibnamefont{Zhang}}, \bibnamefont{and}
  \bibinfo{author}{\bibfnamefont{Y.}~\bibnamefont{Qian}},
  {``}\bibinfo{title}{Isostructural Cd$_3$E$_2$ (E = P, As) microcrystals
  prepared via a hydrothermal route},{''} \bibinfo{journal}{Crystal Growth \&
  Design} \textbf{\bibinfo{volume}{6}}, \bibinfo{pages}{849}
  (\bibinfo{year}{2006}).

\bibitem[{\citenamefont{Cheng et~al.}(2016)\citenamefont{Cheng, Zhang, Liu,
  Yuan, Song, Sun, Zhou, Zhang, and Xiu}}]{Cheng2016NJP}
\bibinfo{author}{\bibfnamefont{P.}~\bibnamefont{Cheng}},
  \bibinfo{author}{\bibfnamefont{C.}~\bibnamefont{Zhang}},
  \bibinfo{author}{\bibfnamefont{Y.}~\bibnamefont{Liu}},
  \bibinfo{author}{\bibfnamefont{X.}~\bibnamefont{Yuan}},
  \bibinfo{author}{\bibfnamefont{F.}~\bibnamefont{Song}},
  \bibinfo{author}{\bibfnamefont{Q.}~\bibnamefont{Sun}},
  \bibinfo{author}{\bibfnamefont{P.}~\bibnamefont{Zhou}},
  \bibinfo{author}{\bibfnamefont{D.~W.} \bibnamefont{Zhang}}, \bibnamefont{and}
  \bibinfo{author}{\bibfnamefont{F.}~\bibnamefont{Xiu}},
  {``}\bibinfo{title}{Thickness-dependent quantum oscillations in Cd$_3$As$_2$
  thin films},{''} \bibinfo{journal}{New J. Phys.}
  \textbf{\bibinfo{volume}{18}}, \bibinfo{pages}{083003}
  (\bibinfo{year}{2016}).

\bibitem[{\citenamefont{Clement}(1953)}]{Clem1953}
\bibinfo{author}{\bibfnamefont{J.~R.} \bibnamefont{Clement}},
  {``}\bibinfo{title}{The atomic heat and critical magnetic field of
  superconducting cadmium},{''} \bibinfo{journal}{Phys. Rev.}
  \textbf{\bibinfo{volume}{92}}, \bibinfo{pages}{1578} (\bibinfo{year}{1953}).

\bibitem[{\citenamefont{Goodman and Mendoza}(1951)}]{good1951}
\bibinfo{author}{\bibfnamefont{B.~B.} \bibnamefont{Goodman}} \bibnamefont{and}
  \bibinfo{author}{\bibfnamefont{E.}~\bibnamefont{Mendoza}},
  {``}\bibinfo{title}{LXII. The critical magnetic fields of aluminium, cadmium,
  gallium and zinc},{''} \bibinfo{journal}{Phil. Mag.}
  \textbf{\bibinfo{volume}{42}}, \bibinfo{pages}{594} (\bibinfo{year}{1951}).

\bibitem[{\citenamefont{Yan et~al.}(2018)\citenamefont{Yan, Khalsa, Vishwanath,
  Han, Wright, Rouvimov, Katzer, Nepal, Downey, Muller et~al.}}]{Yan2018}
\bibinfo{author}{\bibfnamefont{R.}~\bibnamefont{Yan}},
  \bibinfo{author}{\bibfnamefont{G.}~\bibnamefont{Khalsa}},
  \bibinfo{author}{\bibfnamefont{S.}~\bibnamefont{Vishwanath}},
  \bibinfo{author}{\bibfnamefont{Y.}~\bibnamefont{Han}},
  \bibinfo{author}{\bibfnamefont{J.}~\bibnamefont{Wright}},
  \bibinfo{author}{\bibfnamefont{S.}~\bibnamefont{Rouvimov}},
  \bibinfo{author}{\bibfnamefont{D.~S.} \bibnamefont{Katzer}},
  \bibinfo{author}{\bibfnamefont{N.}~\bibnamefont{Nepal}},
  \bibinfo{author}{\bibfnamefont{B.~P.} \bibnamefont{Downey}},
  \bibinfo{author}{\bibfnamefont{D.~A.} \bibnamefont{Muller}},
  \bibnamefont{et~al.}, {``}\bibinfo{title}{GaN/NbN epitaxial
  semiconductor/superconductor heterostructures},{''} \bibinfo{journal}{Nature}
  \textbf{\bibinfo{volume}{555}}, \bibinfo{pages}{183} (\bibinfo{year}{2018}).

\bibitem[{\citenamefont{Chopra}(1979)}]{chopra1979}
\bibinfo{author}{\bibfnamefont{K.~L.} \bibnamefont{Chopra}},
  \emph{\bibinfo{title}{Thin film phenomena}} (\bibinfo{publisher}{Malabar :
  Robert E. Krieger Publishing Company}, \bibinfo{year}{1979}), ISBN
  \bibinfo{isbn}{9780882757469}.

\bibitem[{\citenamefont{Shapoval}(1966)}]{Shapoval1966JETP}
\bibinfo{author}{\bibfnamefont{E.~A.} \bibnamefont{Shapoval}},
  {``}\bibinfo{title}{Critical fields of thin superconducting films},{''}
  \bibinfo{journal}{Sovi. Phis. JETP} \textbf{\bibinfo{volume}{22}},
  \bibinfo{pages}{647} (\bibinfo{year}{1966}).

\bibitem[{\citenamefont{Kuchinskii et~al.}(2017)\citenamefont{Kuchinskii,
  Kuleeva, and Sadovskii}}]{Kuch2017JETP}
\bibinfo{author}{\bibfnamefont{E.~Z.} \bibnamefont{Kuchinskii}},
  \bibinfo{author}{\bibfnamefont{N.~A.} \bibnamefont{Kuleeva}},
  \bibnamefont{and} \bibinfo{author}{\bibfnamefont{M.~V.}
  \bibnamefont{Sadovskii}}, {``}\bibinfo{title}{Temperature dependence of the
  upper critical field in disordered Hubbard model with attraction},{''}
  \bibinfo{journal}{J. Exp. Theor. Phys.} \textbf{\bibinfo{volume}{125}},
  \bibinfo{pages}{1127} (\bibinfo{year}{2017}).

\bibitem[{\citenamefont{Kirshenbaum et~al.}(2013)\citenamefont{Kirshenbaum,
  Syers, Hope, Butch, Jeffries, Weir, Hamlin, Maple, Vohra, and
  Paglione}}]{kirsh2013}
\bibinfo{author}{\bibfnamefont{K.}~\bibnamefont{Kirshenbaum}},
  \bibinfo{author}{\bibfnamefont{P.~S.} \bibnamefont{Syers}},
  \bibinfo{author}{\bibfnamefont{A.~P.} \bibnamefont{Hope}},
  \bibinfo{author}{\bibfnamefont{N.~P.} \bibnamefont{Butch}},
  \bibinfo{author}{\bibfnamefont{J.~R.} \bibnamefont{Jeffries}},
  \bibinfo{author}{\bibfnamefont{S.~T.} \bibnamefont{Weir}},
  \bibinfo{author}{\bibfnamefont{J.~J.} \bibnamefont{Hamlin}},
  \bibinfo{author}{\bibfnamefont{M.~B.} \bibnamefont{Maple}},
  \bibinfo{author}{\bibfnamefont{Y.~K.} \bibnamefont{Vohra}}, \bibnamefont{and}
  \bibinfo{author}{\bibfnamefont{J.}~\bibnamefont{Paglione}},
  {``}\bibinfo{title}{Pressure-Induced Unconventional Superconducting Phase in
  the Topological Insulator ${\mathrm{Bi}}_{2}{\mathrm{Se}}_{3}$},{''}
  \bibinfo{journal}{Phys. Rev. Lett.} \textbf{\bibinfo{volume}{111}},
  \bibinfo{pages}{087001} (\bibinfo{year}{2013}).

\bibitem[{\citenamefont{Kong et~al.}(2013)\citenamefont{Kong, Zhang, Zhang,
  Zhu, Liu, Yu, Fang, Jin, Yang, Yu et~al.}}]{kong2013}
\bibinfo{author}{\bibfnamefont{P.~P.} \bibnamefont{Kong}},
  \bibinfo{author}{\bibfnamefont{J.~L.} \bibnamefont{Zhang}},
  \bibinfo{author}{\bibfnamefont{S.~J.} \bibnamefont{Zhang}},
  \bibinfo{author}{\bibfnamefont{J.}~\bibnamefont{Zhu}},
  \bibinfo{author}{\bibfnamefont{Q.~Q.} \bibnamefont{Liu}},
  \bibinfo{author}{\bibfnamefont{R.~C.} \bibnamefont{Yu}},
  \bibinfo{author}{\bibfnamefont{Z.}~\bibnamefont{Fang}},
  \bibinfo{author}{\bibfnamefont{C.~Q.} \bibnamefont{Jin}},
  \bibinfo{author}{\bibfnamefont{W.~G.} \bibnamefont{Yang}},
  \bibinfo{author}{\bibfnamefont{X.~H.} \bibnamefont{Yu}},
  \bibnamefont{et~al.}, {``}\bibinfo{title}{Superconductivity of the
  topological insulator ${\mathrm{Bi}}_{2}{\mathrm{Se}}_{3}$ at high
  pressure},{''} \bibinfo{journal}{J. Phys.: Condens. Matter}
  \textbf{\bibinfo{volume}{25}}, \bibinfo{pages}{362204}
  (\bibinfo{year}{2013}).

\bibitem[{\citenamefont{Nishihaya et~al.}(2018)\citenamefont{Nishihaya, Uchida,
  Nakazawa, Akiba, Kriener, Kozuka, Miyake, Taguchi, Tokunaga, and
  Kawasaki}}]{Nish2018PRB}
\bibinfo{author}{\bibfnamefont{S.}~\bibnamefont{Nishihaya}},
  \bibinfo{author}{\bibfnamefont{M.}~\bibnamefont{Uchida}},
  \bibinfo{author}{\bibfnamefont{Y.}~\bibnamefont{Nakazawa}},
  \bibinfo{author}{\bibfnamefont{K.}~\bibnamefont{Akiba}},
  \bibinfo{author}{\bibfnamefont{M.}~\bibnamefont{Kriener}},
  \bibinfo{author}{\bibfnamefont{Y.}~\bibnamefont{Kozuka}},
  \bibinfo{author}{\bibfnamefont{A.}~\bibnamefont{Miyake}},
  \bibinfo{author}{\bibfnamefont{Y.}~\bibnamefont{Taguchi}},
  \bibinfo{author}{\bibfnamefont{M.}~\bibnamefont{Tokunaga}}, \bibnamefont{and}
  \bibinfo{author}{\bibfnamefont{M.}~\bibnamefont{Kawasaki}},
  {``}\bibinfo{title}{Negative magnetoresistance suppressed through a
  topological phase transition in
  $({\mathrm{Cd}}_{1\ensuremath{-}x}{\mathrm{Zn}}_{x}{)}_{3}{\mathrm{As}}_{2}$
  thin films},{''} \bibinfo{journal}{Phys. Rev. B}
  \textbf{\bibinfo{volume}{97}}, \bibinfo{pages}{245103}
  (\bibinfo{year}{2018}).

\bibitem[{\citenamefont{Pietraszko and Lukaszewicz}(1973)}]{Pietraszko1973}
\bibinfo{author}{\bibfnamefont{A.}~\bibnamefont{Pietraszko}} \bibnamefont{and}
  \bibinfo{author}{\bibfnamefont{K.}~\bibnamefont{Lukaszewicz}},
  {``}\bibinfo{title}{Thermal expansion and phase transitions of Cd$_3$As$_2$
  and Zn$_3$As$_2$},{''} \bibinfo{journal}{Phys.. Stat. Sol. A}
  \textbf{\bibinfo{volume}{18}}, \bibinfo{pages}{723} (\bibinfo{year}{1973}).

\bibitem[{\citenamefont{Grigoriev et~al.}(1996)\citenamefont{Grigoriev,
  Meilikhov, and Radzig}}]{Mejlikhov1991}
\bibinfo{author}{\bibfnamefont{I.~S.} \bibnamefont{Grigoriev}},
  \bibinfo{author}{\bibfnamefont{E.~Z.} \bibnamefont{Meilikhov}},
  \bibnamefont{and} \bibinfo{author}{\bibfnamefont{A.}~\bibnamefont{Radzig}},
  \emph{\bibinfo{title}{Handbook of Physical Quantities}}
  (\bibinfo{publisher}{CRC Press}, \bibinfo{year}{1996}).

\bibitem[{\citenamefont{Novikova}(1974)}]{Novikova1974}
\bibinfo{author}{\bibfnamefont{S.~I.} \bibnamefont{Novikova}},
  \emph{\bibinfo{title}{Thermal expansion of solids}}
  (\bibinfo{publisher}{Nauka}, \bibinfo{year}{1974}).

\bibitem[{\citenamefont{{Shvetsov} et~al.}(2018)\citenamefont{{Shvetsov},
  {Esin}, {Timonina}, {Kolesnikov}, and {Deviatov}}}]{Shvetsov2018arxive}
\bibinfo{author}{\bibfnamefont{O.~O.} \bibnamefont{{Shvetsov}}},
  \bibinfo{author}{\bibfnamefont{V.~D.} \bibnamefont{{Esin}}},
  \bibinfo{author}{\bibfnamefont{A.~V.} \bibnamefont{{Timonina}}},
  \bibinfo{author}{\bibfnamefont{N.~N.} \bibnamefont{{Kolesnikov}}},
  \bibnamefont{and} \bibinfo{author}{\bibfnamefont{E.~V.}
  \bibnamefont{{Deviatov}}}, {``}\bibinfo{title}{{Surface superconductivity in
  a three-dimensional Cd$_3$As$_2$ semimetal}},{''} \bibinfo{journal}{ArXiv
  e-prints}  (\bibinfo{year}{2018}), \eprint{1811.02475}.

\end{thebibliography}

\end{document}